\documentclass[12pt]{article}

\usepackage{amsmath,amssymb}

\newcommand{\ident}{{\rm 1}\!\!\,{\rm l}}

\newcommand{\beq}{\begin{eqnarray}}
\newcommand{\eeq}{\end{eqnarray}}

\begin{document}

\thispagestyle{empty}

\begin{center}


\begin{flushright}
BCCUNY-HEP/06-03
\end{flushright}

\vspace{15pt}
{\large \bf STRING TENSIONS AND REPRESENTATIONS IN ANISOTROPIC
$(2+1)$-DIMENSIONAL WEAKLY-COUPLED YANG-MILLS THEORY  }

\vspace{20pt}

{\bf Peter Orland}$^{\rm a.b.c.}$\!\footnote{giantswing@gursey.baruch.cuny.edu}

\vspace{8pt}

\begin{flushleft}
a. Physics Program, The Graduate School and University Center,
The City University of New York, 365 Fifth Avenue,
New York, NY 10016, U.S.A.
\end{flushleft}

\begin{flushleft}
b. Department of Natural Sciences, Baruch College, The 
City University of New York, 17 Lexington Avenue, New 
York, NY 10010, U.S.A. 
\end{flushleft}

\begin{flushleft}
c. The Niels Bohr Institute, Blegdamsvej 17, DK-2100, Copenhagen {\O}, Denmark
 \end{flushleft}

\vspace{40pt}

{\bf Abstract}
\end{center}

\noindent 
In earlier papers we established quark 
confinement analytically in anisotropic $(2+1)$-dimensional 
Yang-Mills theory 
with two gauge
coupling constants. Here we point out a few features of the confining phase. These are: 1) the
string tension in the $x^{2}$-direction as a function of representation obeys a sine law, and 2) static
adjoint sources are not confined.
\hfill

\newpage

\setcounter{footnote}{0}

\setcounter{page}{1}

\section{Introduction}
\setcounter{equation}{0}
\renewcommand{\theequation}{1.\arabic{equation}}

Recently it was found that in anisotropic version of $(2+1)$-dimensional Hamiltonian SU($N$) gauge theory, confinement can be understood analytically at weak coupling \cite{PhysRevD71}, \cite{hep-th/0607013}. The anisotropy is due to the coefficient of one of the electric-field-squared terms in the Hamiltonian
being much smaller than the other. This means that there are two coupling 
constants in the model, both
of which are small. 

The method exploits the fact that in an axial gauge, the Hamiltonian lattice SU($N$) gauge
theory is a set of ${\rm SU}(N)\!\times\!{\rm SU}(N)$ principal chiral models, coupled together
by a nonlocal term. The mass spectra of these models are precisely 
known \cite{Zamolodchikov}, \cite{abda-wieg}, \cite{pol-wieg}. In reference \cite{PhysRevD71}, this information was used to show that confinement
occurs in the anisotropic limit. The string tension in the $x^{2}$-direction was found, and that
in the $x^{1}$-direction was estimated more crudely. In the second paper  \cite{hep-th/0607013},
the string tension in the $x^{1}$-direction for $N=2$ was found more precisely using the exact form
factors of current operators of the principal chiral model \cite{KarowskiWeisz}. 

For other recent analytic approaches to confinement in $(2+1)$ dimensions, see
references
\cite{kar-nair}, \cite{orl-sem}, \cite{orl-gauge-inv}, \cite{leigh-min-yel}.

In this paper, we point out a few simple consequences of this mechanism of confinement. In
particular we consider charges other than elementary quarks. We point out that static sources built from $k$ elementary colors are confined
with a string tension proportional to $\sin\pi k/N$ if they are separated in the $x^{2}$-direction. The dependence of the string tension on $k$ in the $x^{1}$-direction in more subtle (as system
is anisotropic, the lattice is not invariant under $90^o$ rotations). The dependence on $k$ in the 
$x^{1}$-direction has a leading term proportional
to the quadratic Casimir  \cite{PhysRevD71}. There are, however, nontrivial 
corrections, which, as
mentioned above, have only been
computed for the case of gauge group SU($2$) \cite{hep-th/0607013}. We also find that adjoint sources
are not confined.

The dependence of the string tension between two sources as a function of the representation
is somewhat controversial. Lattice strong-coupling expansions \cite{KogutSusskind} and 
dimensional-reduction
arguments \cite{AmbjornOlesenPeterson} imply a Casimir law. Further support for
this result and some implications for confinement mechanisms were put
forward \cite{Greensite}. In contrast, a sine law 
was found in ${\mathcal N}=2$ supersymmetric gauge theory softly broken to ${\mathcal N}=1$
\cite{DouglasShenker}, in so-called MQCD \cite{HananyStrasslerZaffaroni},
and in the AdS/QCD scheme \cite{HerzogKlebanov}. Most numerical studies favor
a result midway between the Casimir law and the sine law \cite{DelDebbio1}, \cite{Lucini1}, 
\cite{Lucini2}, though one large-scale calculation
favors the latter \cite{DelDebbio2}. A general picture of the sine law was proposed by Gliozzi
\cite{Gliozzi}. A summary of the situation and the general issues involved can be found
in Section 4.2 of reference \cite{GreensiteReview}.

In the next section, we review the anisotropic lattice Hamiltonian in the axial gauge. We explain
the sine law behavior of the string tensions in the $x^{2}$ direction in Section 3. We touch briefly
on the $k$ dependence of the string tension in the $x^{1}$ direction in Section 4. It is explained
why adjoint sources are not confined in Section 5. In the last section, we summarize our results
and mention some problems for the future.

\section{The Anisotropic Regularized Hamiltonian}
\setcounter{equation}{0}
\renewcommand{\theequation}{2.\arabic{equation}}

We briefly review the axial-gauge formulation of the Hamiltonian lattice gauge theory. A more detailed discussion of the 
axial-gauge fixing procedure can be found in references
\cite{PhysRevD71}, \cite{hep-th/0607013}.

Space is taken to be a cylinder
of dimensions $L^{1}\times L^{2}$, with open boundary conditions in the $x^{1}$-direction and
periodic boundary conditions in the $x^{2}$-direction \cite{PhysRevD71}. The range
of coordinates is $x^{1}=0,a,2a,\dots,L^{1}$, $x^{2}=0,a,2a,\dots, L^{2}-a$. For any function 
$f(x^{1},x^{2})$, we require that $f(x^{1},x^{2}+L^{2})=f(x^{1},x^{2})$.  We impose no conditions
on $f(0,x^{2)}$ and $f(L^{1},x^{2})$, at the ends of the cylinder. The lattice spacing is
denoted by $a$. The generators  of 
SU($N$), $t_{b}$,  $b=1,\dots, N^{2}-1$, satisfy ${\rm Tr}\,t_{b}t_{c}=\delta_{bc}$, and
$[t_{b},t_{c}]={\rm i}f_{bc}^{d}t_{d}$. We begin with
with two gauge fields on the links of the lattice $U_{1}(x^{1},x^{2}), U_{2}(x^{1},x^{2})\in {\rm SU}(N)$
and with electric-field operators
$l_{1}(x^{1},x^{2})_{b}$, $l_{2}(x^{1},x^{2})_{b}$, $b=1,\dots,N^{2}-1$, satisfying the commutation
relations
\beq
[l_{j}(x)_{b},U_{k}(y)]=-\delta_{jk}\delta_{xy}t_{b}U_{j}(x)\;,\;\;
[l_{l}(x)_{b},l_{k}(y)_{c}]={\rm i}\delta_{jk}\delta_{xy}f_{bc}^{d} l_{j}(x)_{d}\;,
\nonumber
\eeq
with all other commutators zero. We denote 
the adjoint representation of the SU($N$) gauge field by 
$\mathcal R_{j}(x)$. The precise definition is
${\mathcal R}_{j}(x)_{b}^{\;\;c}t_{c}=
U_{j}(x)t_{b}U_{j}(x)^{\dagger}$.  

We can introduce fundamental source operators $q(x)_{b}$, $b=1,\dots,N^{2}-1$ at sites, which must obey the commutation
relations
\begin{eqnarray}
[q(x)_{b},q(y)_{c}]={\rm i} f_{bc}^{d} \delta_{xy}q(x)_{d} \; . \label{charge-comm}
\end{eqnarray}
The temporal gauge $A_{0}=0$ is taken. For any physical
state $\Psi$, Gauss's law may be written as
\beq
l_{1}(x^{1},x^{2})\Psi =\left[{\mathcal R}_{1}(x^{1}-a,x^{2})l_{1}(x^{1}-a,x^{2})-
({\mathcal D}_{2}\, l_{2}) (x^{1}-a, x^{2}) +q(x) \right] \Psi \;,  \nonumber
\eeq
where ${\mathcal D}_{2}\,l_{2}) (y^{1}, x^{2})=U(y^{1},x^{2})-{\mathcal R}_{2}(y^{1},x^{2}-a)
U_{2}(y^{1},x^{2}-a)$. Thus the operators on each side of this expression may be identified. Taking
the gauge condition $U_{1}(x^{1},x^{2})=\ident$, which is possible everywhere on
a cylinder, we sum over the $1$-coordinate yielding
\beq
l_{1}(x^{1},x^{2})=
\sum_{y^{1}=0}^{x^{1}}
q(y^{1},x^{2})
-\sum_{y^{1}=0}^{x^{1}}({\mathcal D}_{2}\,l_{2}) (y^{1}, x^{2}) \;.
\label{lattice-electric}
\eeq

Gauss's law is not fully imposed by (\ref{lattice-electric}), which must be supplemented
by the condition
\beq
\left[
\sum_{x^{1}=0}^{L^{1}}({\mathcal D}_{2}\,l_{2})(x^{1},x^{2})-\sum_{x^{1}=0}^{L^{1}}q(x^{1},x^{2})
\right] \Psi=0 \label{lattice-physical}
\eeq
on physical states $\Psi$. Such expressions were considered in the continuum by 
Mandelstam \cite{Mandelstam}.

In the axial gauge, using the nonlocal expression (\ref{lattice-electric})
for the electric field in the $x^{1}$-direction
(which henceforth will be called the horizontal direction), the Hamiltonian 
is $H=H_{0}+H_{1}$, where 
\beq
H_{0}=\sum_{x^{2}=0}^{L^{2}-a} H_{0}(x^{2})\;, \label{sum-of-two-terms}
\eeq
with
\beq
H_{0}(x^{2})= \sum_{x^{1}=0 }^{L^{1}} 
\;\frac{g_{0}^{2}}{2a}[ l_{2}(x^{1},x^{2})]^{2}-
\sum_{x=0}^{L^{1}-a} 
\frac{1}{2g_{0}^{2}a}\;
\left[ {\rm Tr}\; U_{2}(x)^{\dagger}U_{2}(x^{1}+a,x^{2}) +c.c.\right] \;, \label{lattice-sigma}
\end{eqnarray}
and
\beq
H_{1}&=& \frac{(g_{0}^{\prime})^{2}}{2a}\;\sum_{x^{1}=0 }^{L^{1}} \sum_{x^{2}=0}^{L^{2}-a}
[l_{1}(x^{1},x^{2})]^{2}  \nonumber \\
&=&-\frac{(g_{0}^{\prime})^{2}}{2a}\;\sum_{x^{2}=0}^{L^{2}-a} \;\sum_{x^{1},y^{1}=0}^{L^{1}}
{\vert x^{1}-y^{1}\vert } 
\left[{\mathcal D}_{2} l_{2}(x^{1},x^{2})-q(x^{1},x^{2})   \right]   \nonumber  \\
&\times& \!\!\left[ {\mathcal D}_{2}l_{2}(y^{1},x^{2})
-q(y^{1},x^{2})    \right] \;,
\label{nonlocal}
\eeq
and where we have now introduced the second dimensionless coupling 
constant $g_{0}^{\prime}$. The only
interaction in the $x^{2}$-direction (which from now on will be called the vertical direction)
is due to $H_{1}$.

The operators $H_{0}(x^{2})$ are Hamiltonians of ${\rm SU}(N)\!\times\!{\rm SU}(N)$ principal chiral nonlinear
sigma models with coupling constant $g_{0}$ \cite{PhysRevD71},  \cite{hep-th/0607013}. These
are regularized versions of the system with the Lagrangian 
\beq
{\mathcal L}=\frac{1}{2g_{0}^{2}}\eta^{\mu \nu}{\rm Tr}\partial_{\mu} U^{\dagger} \partial_{\nu}U \;,
\;\;\mu,\nu=0,1\;.\nonumber
\eeq
where we identify $U$ with $U_{2}$. Each 
sigma model is regulated on a horizontal ladder of plaquettes. There
is a such a ladder for each value of $x^{2}$. Each ladder can be thought of as a 
layer in which a sigma model is present. Setting $g_{0}^{\prime}$ to zero results
in decoupled layers of $(1+1)$-dimensional sigma models. Increasing  $g_{0}^{\prime}$
leads to an interaction between the vertically-separated layers. This
fact was used to give a set of 
simple arguments for confinement for $g_{0}^{\prime}\ll g_{0}$. The expression for $H_{1}$, acting
on physical states, with $q=0$, is 
identical to the expression on the right-hand side of equation (3.6) in reference
\cite{PhysRevD71}, by virtue of (\ref{lattice-physical}).

The Hamiltonians $H_{0}(x^{2})$ have particle excitations with a non-trivial
S-matrix \cite{abda-wieg}. There is a fundamental particle, with mass gap $m_{1}$. Other
particles are bound
states of $k$ of these fundamental particles. We call all of these excitations for $k=1,\dots,N-1$
Faddeev-Zamolodchikov or FZ particles. The mass spectrum as a function of  
$k$ is
\beq
m_{k}=m_{1}\frac{\sin\frac{\pi k}{N}}{\sin\frac{\pi}{N}} \;, \;\; 
m_{1}=\frac{C_{N}}{g_{0}{\sqrt{N}} a}e^{-4\pi/(g_{0}^{2}N)}+\cdots\;,
\label{mass-formula}
\eeq
where the dots denote corrections and $C_{N}$ is a constant \cite{abda-wieg}. Both the corrections and the value
of $C_{N}$ are non-universal. Perhaps the reader can already see the origin of the sine
law for the string tension in the $x^{2}$-direction, henceforth called the vertical
string tension, in equation (\ref{mass-formula}). If not, it will be explained in the next 
section.

The nonlocal form of the interaction $H_{1}$ can be made local by reintroducing the
temporal gauge field $\Phi=A_{0}$. In one continuous infinite dimension
$\mathbb R$, the function
$g(x^{1}-y^{1})=\vert x^{1}-y^{1} \vert/2$ is the Green's function of the second-derivative
operator. On the lattice, with $x^{1}$ and $y^{1}$ taking values $0, \;a,\;2a,\;\dots,\;L^{1}$, the
same function $g(x^{1}-y^{1})=\vert x^{1}-y^{1} \vert/2$ is the Green's function of an $(L^{1}/a+1)$-dimensional operator
$\Delta_{L^{1},a}$, by which we mean
\beq
\Delta_{L^{1},a}\,g(x^{1}-y^{1})=
\sum_{z^{1}=0}^{L^{1}}\left( \Delta_{L^{1},a}\right)_{x^{1} z^{1}}\,g(z^{1}-y^{1})=
\frac{1}{a}\delta_{x^{1}y^{1}} \;. \nonumber
\eeq 
In the continuum limit $a\rightarrow 0$ and thermodynamic 
limit $L^{1}\rightarrow \infty$, $\Delta_{L^{1},a}\rightarrow -\partial_{1}^{2}$. We can introduce
an auxiliary field $\Phi(x^{1},x^{2})_{b}$ to replace (\ref{nonlocal}) by
\beq
H_{1}&=& \sum_{x^{2}=0}^{L^{2}-a}  \sum_{x^{1}=0}^{L^{2}} \left\{
\frac{(g_{0}^{\prime})^{2}a}{4}\,\Phi(x^{1},x^{2})\Delta_{L^{1},a}\Phi(x^{1},x^{2}) \right. \nonumber \\
&-\!\!&\!\! (g_{0}^{\prime})^{2}
\left[ l_{2}(x^{1},x^{2})-{\mathcal R}_{2}(x^{1},x^{2}-a)l_{2}(x^{1},x^{2}-a) \right. \nonumber \\ 
&-&q(x^{1},x^{2})   \left]  
\Phi(x^{1},x^{2})  \right\} .
\label{local}
\eeq

\section{Vertical $k$-String Tensions}
\setcounter{equation}{0}
\renewcommand{\theequation}{3.\arabic{equation}}

In reference \cite{PhysRevD71} a simple explanation of quark confinement was given
for $g_{0}^{\prime}\ll g_{0}$. The general picture of glue excitations
was discussed in reference \cite{hep-th/0607013}. 

Let us first suppose that no external charges are present in the gauge theory. If $g_{0}^{\prime}=0$, the lattice splits into layers, each layer containing a principal chiral sigma model. The
layers are labeled by the coordinate $x^{2}$. The FZ excitations in each layer may move 
in the $x^{1}$ direction, but may not move between layers. The motion of these particles in a layer
is not free, as particles
scatter one another and form
bound states; this dynamics is governed by the S-matrix given in reference \cite{abda-wieg}. There is no
particle production in this integrable dynamics.

If we now increase $g_{0}^{\prime}$
slightly, two new types of interaction are present. First,  a linear
potential is produced between FZ particles
in the same layer and in neighboring layers. Second, there will be the possibility of particle
creation and destruction, the dynamics being no longer integrable.

Suppose a source constructed from $k$ fundamental static quarks lies at
$x^{2}=u^{2}$ and a source constructed from $k$ fundamental anti-quarks lies at 
$x^{2}=v^{2}>u^{2}$. The
lowest-energy state consistent with the
residual Gauss's law condition (\ref{lattice-physical}) is made as follows. Each layer with $x^{2}<u^{2}$
and with $x^{2}\ge v^{2}$ is in the vacuum state. The layers with $u^{2}\le x^{2}<v^{2}$ must contain
$k$ fundamental FZ particles. The lowest energy is produced by binding these $k$ particles
in a given layer into a single bound state. The binding energy is 
\beq
E_{\rm binding}=\left(k-\frac{\sin \frac{\pi k}{N}}{\sin\frac{\pi}{N}}\right)m_{1} >0\;. \nonumber
\eeq
The vertical $k$-string tension is just the energy of the bound state of $k$ fundamental
FZ particles, divided by the lattice spacing:
\beq
\sigma_{\rm V}^{k}=\frac{m_{k}}{a}
= \frac{C_{N}}{g_{0}{\sqrt{N}}a^{2}} \,\frac{\sin \frac{\pi k}{N}}{\sin\frac{\pi}{N}}\, e^{-4\pi/(g_{0}^{2}N)}
+\cdots\;,
\label{vertical-string-tensions}
\eeq
where the dots indicate non-universal corrections.

We could worry that there could be a drastic change in the behavior of the 
string tension for $g_{0}^{\prime}>0$, i.e. that perturbation theory in (2.6)
could break down. There are several reasons why we believe this does not
occur and that
the formula (\ref{vertical-string-tensions}) should be approximately correct
as $g_{0}^{\prime}\ll g_{0}$. 

The
term (2.6) only binds FZ particles horizontally. If $g_{0}^{\prime}=0$, the vertical string
tension is simply the mass $m_{k}$ divided by the lattice spacing. Increasing
$g_{0}^{\prime}$ slightly will lead to a binding between FZ particles in adjacent layers, but
have no other effect. For sufficiently large $g_{0}^{\prime}/g_{0}$ other effects will become
important, most notably dielectric breakdown; horizontal electric flux will become unstable to the
production of FZ particles. Dielectric breakdown will occur only for sufficiently heavy horizontal
electric flux. For sufficiently small $g_{0}^{\prime}$, there should be no dielectric breakdown.

Further evidence that an expansion in $g_{0}^{\prime}$ is reliable was found
in reference \cite{hep-th/0607013}. There the horizontal string tension for $N=2$ was
worked out in perturbation theory in $g_{0}^{\prime}$. The leading term of the result is 
consistent with the results of this paper. This indicates that perturbation theory in
$g_{0}^{\prime}$ works very well, at least for $g_{0}^{\prime}\ll g_{0}$.

\section{Horizontal $k$-string tensions}
\setcounter{equation}{0}
\renewcommand{\theequation}{4.\arabic{equation}}

We now briefly remark on horizontal string tensions, by which we mean the string tensions of sources
separated in the $x^{1}$-direction.

Suppose that a source made from $k$ static fundamental quarks is placed at $x^{1}=u^{1}$, 
$x^{2}=u^{2}$ and a source made from $k$ static fundamental anti-quarks is placed at $x^{1}=v^{1}$, 
$x^{2}=u^{2}$. Such a configuration is consistent with the condition (\ref{lattice-physical}) 
on physical states. If the coupling $g_{0}^{\prime}$ is sufficiently small, then the existence
of the mass gap in the principal chiral sigma models at $x^{2}=u^{2}$ and $x^{2}=u^{2}-a$ forces
electric flux along the line from $(u^{1},u^{2})$ to $(v^{1},u^{2})$ \cite{PhysRevD71}. Thus the
potential is just that of the $(1+1)$-dimensional SU($N$) Yang-Mills theory. This argument
implies that the horizontal string tension should behave as 
\beq
\sigma_{\rm H}^{k}=\left( \frac{g_{0}^{\prime}}{a} \right)^{2}C_{k}\;,
\label{horizontal-string-tensions}
\eeq
where $C_{k}$ is the quadratic Casimir operator. This 
formula as it stands, however, is not sufficient. It is derived on the assumption that the static
potential energy of the two sources receives no 
contributions from the principal chiral sigma models themselves. In fact,
the correlation functions of currents in the sigma models do produce such additional
contributions. Thus far, these
have only been calculated for $k=1$ and gauge group 
SU($2$) \cite{hep-th/0607013}, where the naive result 
(\ref{horizontal-string-tensions}) is reduced to a smaller number.

\section{Adjoint Sources}
\setcounter{equation}{0}
\renewcommand{\theequation}{5.\arabic{equation}}

In this section we show that adjoint sources are 
screened by FZ particles. Consequently, these sources
are not confined.

As preliminary exercise to studying a singlet static adjoint source, consider a quark at
$x^{1}=u^{1}$, $x^{2}=u^{2}$ and an anti-quark at $x^{1}=v^{1}$, $x^{2}=u^{2}+a$. Now project the
states of this pair of color sources to the adjoint representation. All that
is needed is one FZ particle in the sigma model at $x^{2}=u^{2}$ to satisfy the residual gauge-invariance
condition (\ref{lattice-electric}). If the charges are close together $u^{1}\approx v^{1}$, the energy
of the configuration is of order $m_{1}$. We can think of this pair of sources as a ``smeared-out"
adjoint source.

Next, let us consider a point-like adjoint source at one site of the lattice $x^{1}=u^{1},\;x^{2}=u^{2}$. It is
not possible to satisfy (\ref{lattice-electric}) by introducing a single FZ particle, as we did in the 
last paragraph. By introducing
two FZ particles in the layer at $u^{2}$ (or in the layer at $x^{2}-a$), it is possible to make a configuration satisfying (\ref{lattice-electric}). This is a singlet formed from three adjoint particles. The minimum possible energy
is therefore $2m_{1}$, minus the binding energy for these particles. One FZ particle has $k=1$ and the other
has $k=N-1$. Therefore these two FZ particles are not bound to each other
by the sigma-model dynamics. The FZ particles and the adjoint source are
bound, however, by the interaction term $H_{1}$. The smallest possible energy of the pair of FZ 
particles
particles is $2m_{1}$, if we ignore 
the binding energy coming from the interaction. In any case, the
total energy is finite in the infinite-volume limit.

To understand how this works more explicitly, we shall describe the singlet states of three adjoint particles. One
of these is a static source and the other two are FZ particles. 

Let us call the vector space of 
fundamental (quark) states $V_{F}$ and its conjugate ${\bar V}_{F}$. The vector space of an adjoint
particle state $V_{A}$ is a subspace of $V_{F}\otimes{\bar V}_{F}$. There is a projection $P$
to this subspace, $V_{A}=P(V_{F}\otimes{\bar V}_{F})$. If 
$v^{b}{\bar w}^{c}\in V_{F}\otimes {\bar V}_{F}$, the projection to $V_{A}$ is given by
${P}^{bc}_{de} v^{d} {\bar w}^{e}$, where 
\beq
{P}^{bc}_{de}=\delta^{b}_{d}\delta^{c}_{e}-\frac{1}{N}\delta^{bc}\delta_{de}\;. \nonumber
\eeq

The state of three adjoint particles lies in the vector space $V_{A}\otimes V_{A}\otimes V_{A}$. Let
us write this as $P(V_{F}\otimes{\bar V}_{F}) \otimes P(V_{F}\otimes{\bar V}_{F}) \otimes
P(V_{F}\otimes{\bar V}_{F})= P^{3} V_{F}\otimes{\bar V}_{F} \otimes V_{F}\otimes{\bar V}_{F} \otimes
V_{F}\otimes{\bar V}_{F}$, where $P^{3}$ denotes the projection of all of the three $P$'s. Suppose
that $v^{b}{\bar w}^{c}u^{d}{\bar x}^{e} t^{f} {\bar s}^{g}$ is a vector in
the three fundamental, three anti-fundamental vector space $V_{F}\otimes{\bar V}_{F} \otimes V_{F}\otimes{\bar V}_{F} \otimes
V_{F}\otimes{\bar V}_{F}$, there are two projections to singlets. We denote these as
$Q^{bcdefg}_{hijklm}v^{h}{\bar w}^{i}u^{j}{\bar x}^{k} t^{l} {\bar s}^{m}$ and
$R^{bcdefg}_{hijklm}v^{h}{\bar w}^{i}u^{j}{\bar x}^{k} t^{l} {\bar s}^{m}$, where
\beq
Q^{bcdefg}_{hijklm}=\frac{1}{N^{3}}\delta^{cd}\delta^{ef}\delta^{gb}\delta_{ij}\delta_{kl}\delta_{mh}\;,\;\;
R^{bcdefg}_{hijklm}=\frac{1}{N^{3}}\delta^{cf}\delta^{dg}\delta^{be}\delta_{il}\delta_{jm}\delta_{hk}\;.\label{projections}
\eeq
The operators $Q$ and $R$ are projections
from the states of three adjoint particles (which lie in a subspace of
three fundamental 
and three anti-fundamental particles) to singlets. These are the lightest states that can exist
if an adjoint source is present, which are consistent with (\ref{lattice-physical}). Pairs of
Kronecker deltas in (\ref{projections}) correspond to horizontal electric strings tying the
three adjoint particles together. This is how a gauge-invariant finite-energy state with one adjoint
source can exist.

\section{Conclusions}

In this paper we have considered the confining phase of anisotropic $(2+1)$-dimensional SU($N$)
lattice gauge theory. We have
found that the dependence of the string tension on the representation
has a sine law in one spatial
direction and a Casimir law with (significant) corrections in the other spatial direction.

To explicitly calculate the corrections to the sine
law, is a very difficult problem. We are currently attempting to calculate the corrections
to the vertical string tension for $N=2$, using exact form factors. It 
is also desirable to better understand the dependence of the horizontal string tension on $k$. This
calculation could be done
with the knowledge of exact form factors for the ${\rm SU}(N)\! \times \!{\rm SU}(N)$ sigma 
model. However, these
form factors are only known for the $N=2$ case \cite{KarowskiWeisz}.

It would interesting to interpret our results for adjoint sources in the 't\,Hooft limit $N\rightarrow \infty$,
$g_{0}^{2}N$, $(g_{0}^{\prime})^{2}N$ fixed. In this limit, such sources are 
confined \cite{GreensiteHalpern}. The mass $m_{1}$ has a well-defined 't\,Hooft limit. We would
like to know how the simple argument of Section 5 breaks down, when this limit
is taken. We believe that the answer lies in form of the projection operators $Q$ and $R$
which 
project three-adjoint-source (or three-heavy-gluonic) states 
to a singlets. These operators are of order $1/N^{3}$. In the 't\,Hooft limit, the singlet states occupy a
vanishingly small relative volume of state space. They are thus entropically forbidden.

\section*{Acknowledgement}

This 
work was 
supported in part by a grant from the PSC-CUNY.

\end{document}